G. Belokrylov, A. Korenev, B. Lodonova, A. Novokhrestov

# Ensemble of classifiers for speech evaluation

**Abstract**: The article describes an attempt to apply an ensemble of binary classifiers to solve the problem of speech assessment in medicine. A dataset was compiled based on quantitative and expert assessments of syllable pronunciation quality. Quantitative assessments of 7 selected metrics were used as features: dynamic time warp distance, Minkowski distance, correlation coefficient, longest common subsequence (LCSS), edit distance of real sequence (EDR), edit distance with real penalty (ERP), and merge split (MSM). Expert assessment of pronunciation quality was used as a class label: class 1 means high-quality speech, class 0 means distorted. A comparison of training results was carried out for five classification methods: logistic regression (LR), support vector machine (SVM), naive Bayes (NB), decision trees (DT), and K-nearest neighbors (KNN). The results of using the mixture method to build an ensemble of classifiers are also presented. The use of an ensemble for the studied data sets allowed us to slightly increase the classification accuracy compared to the use of individual binary classifiers.

**Keywords:** classifier, classifier ensemble, artificial intelligence, speech assessment, syllable intelligibility, speech rehabilitation.

**Introduction**

Statistics on oncological diseases of the vocal tract [1] show that speech analysis methods in medicine do not lose their relevance today. At the same time, more and more attention is paid not only to methods of treating diseases, but also to methods of restorative medicine, namely speech and voice rehabilitation. In this case, only a slight decrease in the standard of living of patients after treatment is allowed. Rehabilitation measures should be carried out taking into account the individual characteristics of the patient and, if necessary, be adjusted.

Currently, the official (according to clinical recommendations) method of speech assessment is expert speech assessment using a method based on the GOST standard [2]. This method is proposed to be replaced by a method of quantitative automated assessment of the intelligence of syllable pronunciation using distance calculation algorithms and machine learning methods.

Machine learning allows achieving greater efficiency in speech signal analysis by analyzing various user characteristics [3,4]. Therefore, it was proposed to continue research on the application of machine learning methods, namely classification methods. Previously, an evaluation algorithm was proposed based on representing audio signals as a sequence of values, reducing them to the same length using the dynamic time warp (DTW) algorithm, calculating the distance measure between the given sequences and using a fuzzy classifier as a mechanism for combining the calculated values and obtaining the final quantitative assessment. In this paper, we propose an analysis of the approach to constructing an ensemble of binary classifiers for assessing speech intelligibility in comparison with the use of individual binary classifiers.

It is also worth noting that the use of an ensemble of binary classifiers for analyzing "big" data is used in various fields of knowledge, such as medicine, economics, and information security. For example, a combination of 2 or more classifiers can increase the accuracy of detecting DDoS attacks by an average of 5% compared to the accuracy of a single classifier [5].

**Description of data and metrics**

Based on a dataset of audio recordings of patients undergoing treatment and speech rehabilitation at the Tomsk National Research Medical Center oncology research institute [6], quantitative values of similarity and distance metrics were calculated. The set of recordings can be divided into three groups according to three problematic phonemes [k], [s], [t] for the disease in question. Distances were calculated as follows. Each patient had a control session (a set of syllable pronunciation recordings), which is the main one for comparison. The metric value was calculated for the syllable recording in the control and assessed sessions. Also, the recordings in the assessed sessions were assessed by a speech therapist by assigning a class label: 0 - syllable pronunciation in a low-quality recording (inaudible), 1 - syllable pronunciation in a high-quality recording (intelligible). Values were calculated for the following 7 metrics:

1. DTW Distance – Path Cost Estimation in Dynamic Time Warping Algorithm [7];
2. Correlation coefficient;
3. Minkowski distance;
4. Editing distance on real sequences - EDR [8,9];
5. Edit distance on real sequences with penalty - ERP [8,9];
6. Length of the longest common subsequence - Longest Common Subsequence LCSS [10,11];
7. Distance MSM (Move Split Merge) calculation of the number of necessary actions (move, split, merge) to transform one sequence into another [12].

Thus, for each recording in all the sessions being evaluated, a vector of features-values for the metric and a label of the class to which the speech belongs are formed. Based on this label, the considered classifiers and ensembles of classifiers will be trained. For each of the problematic phonemes [k], [s], [t], a set of 1020 feature vectors with a class label was formed.

**Selected classifiers**

Since a binary classification is carried out, the classification methods that are most often used to solve this problem were selected for the study. The following 5 classification methods were studied:

1. KNeighborsClassifier (Knn) - K Nearest Neighbors Method;
2. RandomForestClassifier (RF) - Random Forest Algorithm;
3. Support vector machines (SVC) – Support Vector Machine Classifier;
4. Logistic Regression (LR);
5. DecisionTreeClassifier (DT).

**Rebalancing and cleaning data**

The original data sets are unbalanced, and class 1 prevails in them, it was decided to use methods of noise cleaning and rebalancing of the data sets. For noise cleaning, the quartile analysis algorithm was used. For rebalancing, the oversampling method was chosen using KMeansSMOTE (a combination of two methods for data balancing: K-Means and SMOTE (Synthetic Minority Oversampling Technique)).

Thus, for each of the problematic phonemes, 4 data sets were formed, on which classifiers were subsequently trained.:

1. Original dataset;
2. Dataset with cleaned noise;
3. Rebalanced data set;

4. Rebalanced data set with cleaned noise.

**Method of assembling classifiers**

At this stage, the ensemble method of mixing models Blending was used. The essence of this method is as follows:

1. Several basic models are created;

2. Mixture model training: Base models are trained on a training dataset, and a metamodel is trained on the predictions made by each base model on an independent dataset;

3. After training, a meta_X model emerges, representing the input data that can be used to train the metamodel. Each column or feature represents the output of one base model. Each row represents one sample from an independent dataset.;

4. Then the training of the metamodel begins. It is provided with a classifier, which will be the main one in the training;

5. Based on the combination of results from steps 1-4, the fit_ensemble function is constructed, which trains the mixture model using the training and independent validation datasets.;

6. The mixture ensemble is used to make predictions on new data. It is a two-step process. In the first step, each base model is used to make predictions. The predictions are then pooled together and used as input to the mixture model to make the final prediction. The same cycle was used when training the model. That is, the predictions of each base model were pooled into a training dataset, the predictions were pooled together, and the predict() function was called on the mixture model with this meta-level dataset. The predict_ensemble() function implements these steps. Given a training list of base models, a training ensemble mixer, and a dataset, it returns a set of predictions for a dataset..

7. The get_models() function was then used to create the classification models used in the ensemble. The fit_ensemble() function was then called to train the mixed ensemble on these datasets, and the predict_ensemble() function was used to make predictions on an independent dataset..

8. The effectiveness of the mixing model was assessed by evaluating the classification accuracy..

**Results**

In this study, an attempt was made to apply an ensemble of binary classifiers to solve the problem of speech assessment in medicine. Binary classifiers were trained on three data sets containing quantitative and expert assessments of syllable pronunciation intelligibility. Also, based on these same classifiers, ensembles of classifiers were obtained using the Blending method.:

For the [k] phoneme dataset, the best result on a single classifier was the RandomForestClassifier with an accuracy value of 77.2%. The results were improved when using the Blending ensemble method. The best accuracy result of 78.6% was obtained by mixing the main SVC classifier with additional KNN, SVC, RandomForest and DecisionTree.;

For the [t] phoneme dataset, the best result for a single classifier was 86.3% accuracy on DecisionTree. Using the Bling ensemble method, the results were improved in 24 cases. The highest accuracy result of 87.0% was obtained 5 times, 2 times by mixing the main classifier with 2 additional ones and 3 times by mixing the main classifier with 3 additional ones.;

For the [s] phoneme dataset, the best result for a single classifier was 86.4% accuracy on SVC. Using the Bling ensemble method, the results were confirmed in two cases. The best result of 87% was obtained 2 times: when mixing the main DecisionTree with 3 additional KNN, SVC, LogisticRegression and when mixing the main RandomForest with 3 additional KNN, SVC, LogisticRegression.

On the [s] phoneme, the best result on a single classifier was 86.4 on SVC. Using the ensemble Blenging method, the results were improved in 2 cases. The best result of 87 was obtained 2 times, by mixing the main DecisionTree with 3 additional KNN, SVC, LogisticRegression, and by mixing the main RandomForest with 3 additional KNN, SVC, LogisticRegression.

According to the results of this work, the accuracy results of individual classifiers were slightly improved when using the ensemble mixing method. In the future, other ensemble construction methods will be studied to improve classification accuracy and speech analysis quality assessment.

*Acknowledgments*

This research was funded by the Ministry of Science and Higher Education of Russia, Government Order for 2023–2025, project no. FEWM-2023-0015 (TUSUR).

*References*

1. Malignant neoplasms in Russia in 2021 (Incidence and mortality) [Electronic resource]. – URL: https://www.demoscope.ru/weekly/2023/0975/biblio04.php, (last accessed: 19.09.2024).

2. Standard GOST R 50840-95 Voice over communication paths. Methods for assessing quality, intelligibility and recognition [Electronic resource]. – URL: https://docs.cntd.ru/document/1200027288, (last accessed:: 19.09.2024).

3. In search of a robust facial expression recognition model: A Large-Scale Visual Cross-Corpus Study [Electronic resource]. – URL: https://doi.org/10.1016/j.neucom.2022.10.013, (last accessed: 19.09.2024).

4. Comprehensive paralinguistic analysis of speech: Predicting Gender, Emotions and Deception in a Hierarchical Framework [[Electronic resource]. – URL: https://doi.org/10.21437/Interspeech.2022-11294, (last accessed: 19.09.2024).

5. Detection of DoS attacks using an ensemble of classifiers. Alguliev, Rasim, Alyguliev, Ramiz, Imamverdiev, Yadigar, Sukhostat, Lyudmila. 2017 [Electronic resource]. – URL: https://doi.org/10.25045/NCInfoSec.2017.02, (last accessed: 19.09.2024).

6. Combined measure of similarity of syllable pronunciation based on classification in speech rehabilitation [Electronic resource]. – URL: https://doi.org/10.1007/978-3-031-43792-2_13, (last accessed: 19.09.2024).

7. Time normalization of syllables with the dynamic time warping algorithm in assessing of syllables pronunciation quality when speaking [Electronic resource]. – URL: https://journal.tusur.ru/en/archive/4-2017/time-normalization-of-syllables-with-the-dynamic-time-warping-algorithm-in-assessing-of-syllables-pronunciation-quality-when-speaking, (last accessed: 19.09.2024).

8. Improved algorithm of DTW in speech recognition [Electronic resource]. – URL: https://www.researchgate.net/publication/335082211_Improved_algorithm_of_DTW_in_speech_recognition, (last accessed: 19.09.2024).

9. Improvement of Dynamic Time Warping (DTW) Algorithm [Electronic resource]. – URL: https://www.researchgate.net/publication/300415380_Improvement_of_Dynamic_Time_Warping_DTW_Algorithm, (last accessed: 19.09.2024).

10. Time Series Alignment with Global Invariances [Electronic resource]. –

URL: https://www.researchgate.net/publication/339163589_Time_Series_Alignment_with_Global_Invariances, (last accessed: 19.09.2024).

11. Pattern matching under DTW [Electronic resource]. – URL: https://arxiv.org/abs/2208.14669, (last accessed: 19.09.2024).

12. Comparison of time series similarity measures for plagiarism detection in music [Electronic resource]. – URL: http://ieeexplore.ieee.org/document/7443304, (last accessed: 19.09.2024).